# Feeding Hidden Monsters: a Super-Eddington accreting Black Hole ~1.5 Gyr after the Big Bang

**Hyewon Suh[1], Julia Scharwächter[1], Emanuele Paolo Farina[1], Federica Loiacono[2], Giorgio Lanzuisi[2], Günther Hasinger[3,4,5], Stefano Marchesi[2], Mar Mezcua[6,7], Roberto Decarli[2], Brian C. Lemaux[1,8], Marta Volonteri[9], Francesca Civano[10], Sukyoung K. Yi[11], San Han[11], Mark Rawlings[1], Denise Hung[1]**

[1] Gemini Observatory/NSF's NOIRLab, 670 N. A'ohoku Place, Hilo, HI 96720, USA
[2] INAF - Osservatorio di Astrofisica e Scienza dello Spazio di Bologna, via Gobetti 93/3, 40129, Bologna, Italy
[3] TU Dresden, Institute of Nuclear and Particle Physics, 01062 Dresden, Germany
[4] DESY, Notkestraβe 85, 22607 Hamburg, Germany
[5] Deutsches Zentrum für Astrophysik, Postplatz 1, 02826 Görlitz, Germany
[6] Institute of Space Sciences (ICE, CSIC), Campus UAB, Carrer de Magrans, 08193 Barcelona, Spain
[7] Institut d'Estudis Espacials de Catalunya (IEEC), Edifici RDIT, Campus UPC, 08860 Castelldefels (Barcelona), Spain
[8] Department of Physics and Astronomy, University of California, Davis, One Shields Ave., Davis, CA 95616, USA
[9] Institut d'Astrophysique de Paris (UMR 7095: CNRS & Sorbonne Universite), 98 bis Bd. Arago, F-75014, Paris, France
[10] NASA Goddard Space Flight Center, Greenbelt, MD 20771, USA
[11] Department of Astronomy and Yonsei University Observatory, Yonsei University, Seoul 03722, Republic of Korea

**Recent James Webb Space Telescope (JWST) observations have revealed a surprisingly abundant population of faint, dusty active galactic nuclei (AGNs) at z~4-7[1,2,3,4,5,6,7]. Together with the presence of supermassive black holes (SMBHs) at z>6[8,9,10,11,12,13], this raises questions about the formation and growth histories of early black holes. Current theories[14,15,16] for the formation of seed black holes from the death of the first stars (i.e. light seeds) and/or the direct collapse of primordial gas clouds (i.e. heavy seeds) still lack observational confirmation. Here, we present LID-568, a low-mass ($7.2\times10^6\,M_\odot$) black hole hosting powerful outflows that is observed in an extreme phase of rapid growth at z~4. This object is similar to other JWST-discovered faint AGN populations, but is bright in X-ray emission and accreting at more than 4000% of the limit at which radiation pressure exceeds the force of gravitational attraction of the black hole (i.e. super-Eddington accretion). Analysis of JWST NIRSpec/IFU data reveals spatially extended H$\alpha$ emission with velocities of ~ -600 – -500 km/s relative to the central black hole, indicative of robust

**nuclear-driven outflows. LID-568 represents an elusive low-mass black hole experiencing super-Eddington accretion as invoked by models of early black hole formation. This discovery showcases a previously undiscovered key parameter space and offers crucial insights into rapid black hole growth mechanisms in the early universe.**

Observational surveys have identified several hundreds of luminous quasars at redshift z>6-7 [8,9,10,11,12,13]. The presence of supermassive black holes (SMBHs) with masses of $10^{9-10}\,M_\odot$ at such early cosmic epochs challenges models of SMBH formation and growth, and raises questions about the origin of seed black holes and the mechanisms for their rapid and extreme growth. While the formation of seed black holes remains observationally unconstrained, they are commonly thought to originate in the first galaxies through several gas or stellar physical processes that can generate black holes with masses in excess of $10^2\,M_\odot$ (ref. [17]). Historically, models have been divided into light and heavy seeds, with a demarcation at about $10^3\,M_\odot$. The lightest seeds are generally associated with the death of the first stars with initial masses of $10^{2-3}\,M_\odot$ (ref. [15,16]). The growth of such light seeds at very early time into the observed population of SMBHs at slightly later time is challenging, since black holes formed in this manner would have to accrete at the Eddington limit from the time they are formed up to the redshift at which they are observed[18], which appears to be difficult[19]. The direct collapse of primordial gas clouds into supermassive stars turning into black holes with initial masses of $10^{4-6}\,M_\odot$ (i.e. heavy seed)[20,21] is an attractive alternative, as these heavy seeds can more rapidly grow into SMBHs even via slightly sub-Eddington accretion. However, the expected number densities for the sites where such supermassive stars can form are low. Intermediate pathways where seeds of $10^{3-4}\,M_\odot$ form from very massive stars in pristine rapidly growing halos or through stellar mergers, hierarchical black hole mergers, and stellar captures in dense stellar systems bridge these two extremes[22]. It is also possible that heavy seeds originate from primordial black holes, eliminating the need for the stellar and gas-based processes[23,24,25].

With the unprecedented sensitivity of JWST, it is now possible to extend studies to faint, low-mass sources at high redshifts (i.e. z>3-4), an epoch when both black holes and galaxies are still rapidly growing their mass, and such observation can provide insights into the mechanisms seeding early black holes. JWST has recently discovered a new population of relatively faint, compact, dust-reddened sources at z>4, referred to as "*little red dots*", in a wide variety of extragalactic surveys[1,2,3,4,5,6,7]. These sources are characterized by a red continuum in the rest-frame optical and a modest blue UV continuum. Such sources exhibit prominent broad Balmer emission lines, which imply that they are powered by AGNs. These red compact sources are surprisingly abundant, being 100 times more common than UV-selected quasars at similar redshifts[2]. Furthermore, they are found to have over-massive black holes with respect to the local $M_{BH}$-$M_{stellar}$ relation, exhibiting 10-100 times higher $M_{BH}/M_{stellar}$ ratios[26]. None of these *little red dots* has been detected in X-ray observations[1,2,3,4,5,6,7]. This faint population likely represents the moderate accretion phase of AGN, which are accreting at ~20% of the Eddington rates, and are hosted by relatively low-mass galaxies.

LID-568, an X-ray active galactic nucleus (AGN), was discovered among a hidden black hole population identified as near-infrared (IR)-dropout X-ray sources from the *Chandra*-COSMOS Legacy Survey[27,28]. Similar to *little red dots*, LID-568 appears extremely red and compact in the IR, yet it remains invisible in any optical wavelengths and even in the deepest near-IR imaging taken with the *Hubble Space Telescope (HST)*. Its spectroscopic redshift, $z_{spec}$=3.965, was determined from JWST NIRSpec and MIRI observations, based on broad H$\alpha$, [SII], and Paschen emission lines. The NIRSpec and MIRI spectra of LID-568 are shown in Figure 1. However, LID-568 stands out as uniquely bright in the X-ray relative to the population of *little red dots*, which indicates a higher level of central accretion activity. The observed 0.5-10 keV flux is 5.16 x$10^{-15}$ erg cm$^{-2}$ s$^{-1}$ (ref. [27]). Analysis of the X-ray spectrum (as inferred from the emission measured in the 0.5-2 keV and 2-7 keV bands) allows us to measure the obscuration (log $N_H$=23.44 [-0.34 +0.47] cm$^{-2}$) and the absorption-corrected luminosity in the 0.5-10 keV band (see Methods). The absorption-corrected X-ray luminosity suggests an AGN bolometric luminosity of log $L_{bol}$=46.6 [-0.44 +0.36] erg s$^{-1}$, a factor of ~100 higher than the average bolometric luminosities of *little red dots*.

The shape of the mid- to far-IR spectral energy distribution (SED) of LID-568 exhibits an extremely red IR continuum slope with a single power-law of $\alpha_\lambda \sim$ 4.5 at $\lambda_{rest}\gtrsim$1 µm. This characteristic cannot be explained by the currently available IR SED templates for obscured AGN/ULIRGs and is significantly steeper than those of the *little red dots* (which exhibit $\alpha_\lambda \sim$ 2.0 on average)[4]. The detection of X-ray and mid-IR emission strongly suggests that LID-568 is indeed a heavily obscured AGN, without an apparent presence of the underlying host galaxy features. The model SEDs for super-Eddington accretion suggest a notable absence of rest-frame UV or even optical emission, with a tendency to become progressively redder in the IR as the Eddington ratio increases[29]. However, contrasting perspectives have been presented in other studies, indicating that super-Eddington accretion might lead to an excess of UV radiation, resulting in a significantly bluer continuum slope in the rest-frame UV[30,31].

Given the point-like, compact nature of this source, the extremely red color primarily arises from the thermal emission originating in a dust-obscured accretion disk, with negligible contribution from a host galaxy. Based on IR SED fitting that employs a power-law and two greybodies[32] (as shown in Extended Data Figure 4), the dust temperature is significantly higher (655.53 K and 71.5 K) than what is typically observed in star-forming galaxies (10-60 K). This indicates that hot/warm gas dominates the IR emission, with negligible evidence of star formation activity. This is in contrast to the majority of dust-obscured galaxies at high redshift, which often exhibit signs of powerful starbursts. The derived total IR luminosity is log $L_{8-1000um}$~46.1 erg s$^{-1}$, which is comparable to the AGN bolometric luminosity. The estimated dust mass $M_{dust}$ is ~2.95x$10^6$ $M_\odot$, which suggests that LID-568 contains less dust than the optically-faint, dust-obscured galaxies at

z~3 (i.e. H-dropouts, *HST*-dark, NIR-dark)[33,34] that have dust masses of ~1-4x10$^8$ M$_\odot$. Assuming the dust-to-stellar mass ratios of *HST*-dark, dust-obscured galaxies at similar redshifts[33], the inferred stellar mass of LID-568 is ~2x10$^8$ M$_\odot$, implying a low-mass (i.e. dwarf) galaxy.

The single-epoch virial black hole mass ($M_{BH}$), derived from the broad H$\alpha$ emission line, is 7.2 [-5.4 +10.8]×10$^6$ M$_\odot$, indicating a rather low-mass black hole (see Methods). This yields an Eddington ratio ($L_{bol}/L_{Edd}$) of 41.5, which implies extreme super-Eddington accretion activity. In Figure 2, we show that the black hole mass of LID-568 is comparable to those of faint AGNs discovered by JWST at z~4-7. However, the significantly higher bolometric luminosity of this object places it within a previously unexplored extreme accretion regime. On the other hand, there is growing evidence that AGNs with high accretion rates appear to have smaller broad-line region (BLR) sizes than those predicted by the canonical radius-luminosity relation of sub-Eddington AGNs[35,36]. This discrepancy could potentially lead to an overestimation of the single-epoch black hole mass by as much as ~0.3 dex, resulting in a higher Eddington ratio. LID-568 unveils a missing key parameter space of the extreme accretion and provides critical insights into the rapidly growing mechanisms of black holes for the first time.

The JWST NIRSpec/IFU observation reveals possible evidence of ionized outflows traced by H$\alpha$ emission. In Figure 3, we present NIRSpec/IFU channel maps of the H$\alpha$ emission at different velocity ranges chosen to best highlight the multiple kinematic components observed around the central black hole (see also Extended Data Figure 2). The outflows suggested by the blue-shifted H$\alpha$ emission peak at a projected distance of 0.7" (~5 kpc) to the north-east (A component), 0.4" (~3 kpc) to the north (B component), and at 1" (~7 kpc) towards the south (D component) from the central broad-line component (C component) at ~ -600 – -500 km s$^{-1}$. The continuum emission associated with the spatially extended H$\alpha$ emission components are not detected. The blue-shifted broad [SII] emission is also detected with velocities comparable to those observed in the H$\alpha$ outflows (see Extended data Figure 6).

If the extended H$\alpha$ emissions are associated with outflows, we could infer the AGN lifetime using the outflow velocity and radius. Considering that the outflow reaches ~7 kpc from the central black hole, we obtain the AGN lifetime as t=(7 kpc)/(540 km s$^{-1}$)~1.2x10$^7$ yr. This lifetime is consistent with the lower limits on total accretion timescales set by Soltan arguments (i.e. 10$^{7-9}$ yr; ref. [37]) and indirect measurements of AGN phase timescales (~10$^{7-9}$ yr; ref. [38]). Furthermore, theoretical studies[39] suggest that super-Eddington phases might be sustained over timescales of a few tens of Myr. This lifetime suggests that a substantial fraction of the mass growth of LID-568 may have occurred in a single, super-Eddington accretion episode. However, we note that this could also be associated with stellar feedback-driven outflows from a starburst event preceding the activity in the black hole.

The presence of AGN-driven outflows, along with the lack of significant star-forming activity in LID-568, suggests that AGN feedback may play a crucial role in regulating and/or quenching star formation in this low-mass system at high redshift. This indicates the possibility of rapid and efficient growth of black holes relative to their host galaxies. Theoretical models predict a "blowout" dusty red quasar phase transitioning from a heavily obscured starburst, during which AGN-driven outflows eject gas and dust from the host galaxy, thereby quenching the star formation[40]. It is possible that LID-568 represents a transient phase characterized by extremely high accretion rates with powerful outflows suppressing the star formation in its host galaxy. This could explain the presence of over-massive black holes hosted in low-mass galaxies in the local Universe[41], as well as those found by JWST at z>4[26].

LID-568 could potentially represent the long-sought-after low-mass black hole undergoing rapid growth through super-Eddington accretion. The discovery of a super-Eddington accreting black hole at z~4 offers significant insight regarding the early growth of black holes[42,43,44]. While the rarest, most massive SMBHs at z>6-7 could be explained by an origin from heavy seeds with sub-Eddington accretion, they still require continuous accretion over several hundred million years. The presence of over-massive black hole populations suggests the possibility that they could experience intermittent bursts of super-Eddington growth regardless of whether they originate from heavy or light seeds[44,45]. Super-Eddington accretion likely occurs episodically, and the detection of LID-568 may represent one such episodic accretion phase. Future studies on a large sample of such objects will help constrain the duty cycle of super-Eddington accretion and deepen our understanding of the mechanisms driving such high levels of accretion.

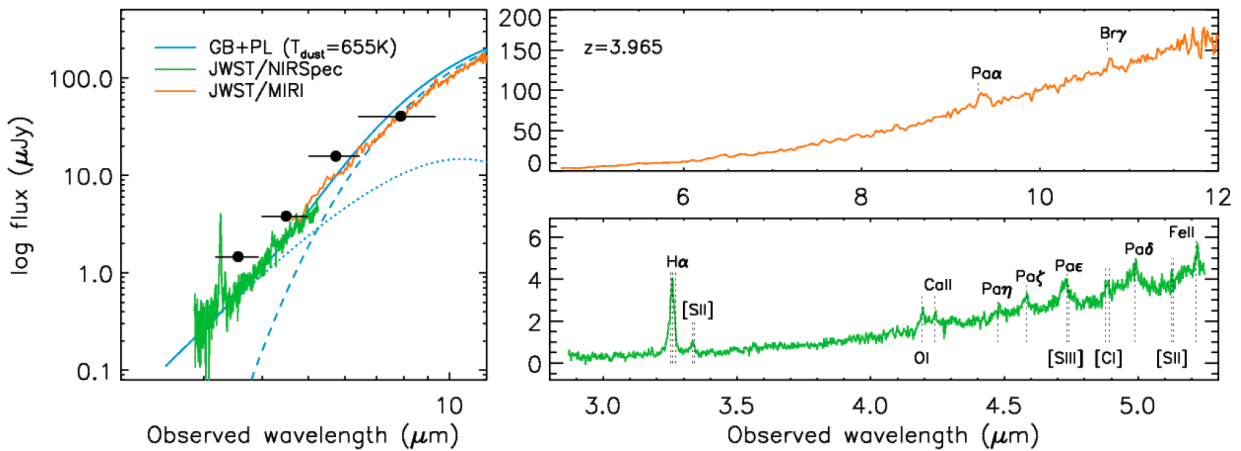

**Figure 1. The NIRSpec and MIRI spectra of LID-568** Left: *Spitzer*/IRAC 3.6, 4.5, 5.8, 8.0 μm photometry (black points) with the best-fitting SED model (blue) at a spectroscopic redshift of $z_{spec}$=3.965 (see Methods). The JWST NIRSpec (green) and MIRI (orange) spectra are overplotted. Right: The spectra of LID-568 obtained with MIRI (top) and NIRSpec (bottom), with the detected emission lines marked.

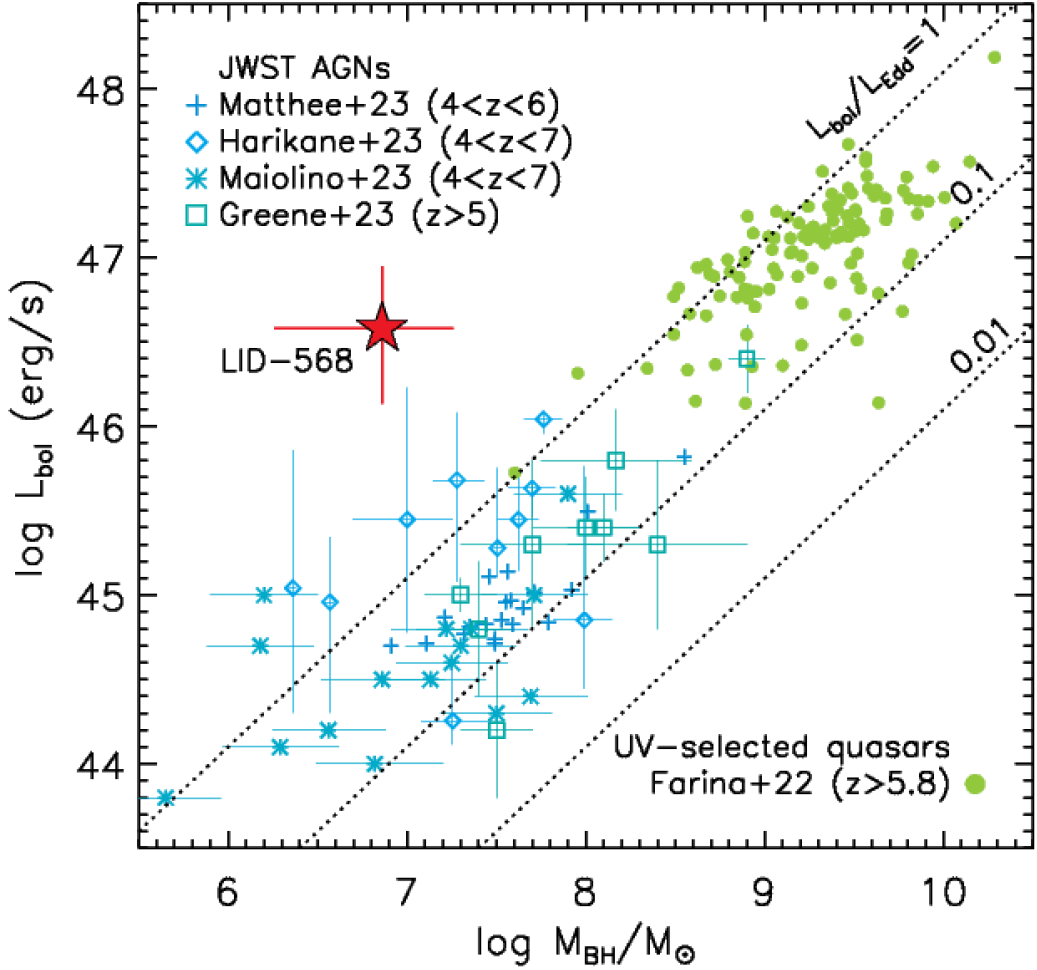

**Figure 2. AGN bolometric luminosity ($L_{bol}$) versus black hole mass ($M_{BH}$) of AGNs at high redshift.** LID-568, with super-Eddington accretion ($L_{bol}/L_{Edd}$~41.5) at z~4, is shown as a red filled star. Its X-ray-derived bolometric luminosity is approximately a factor of ~100 higher than that of faint AGNs at z~4-7 with low-mass black holes[2,3,5,6] recently found by JWST observations. For reference, we also show the UV-selected quasars at z>5.8[12].

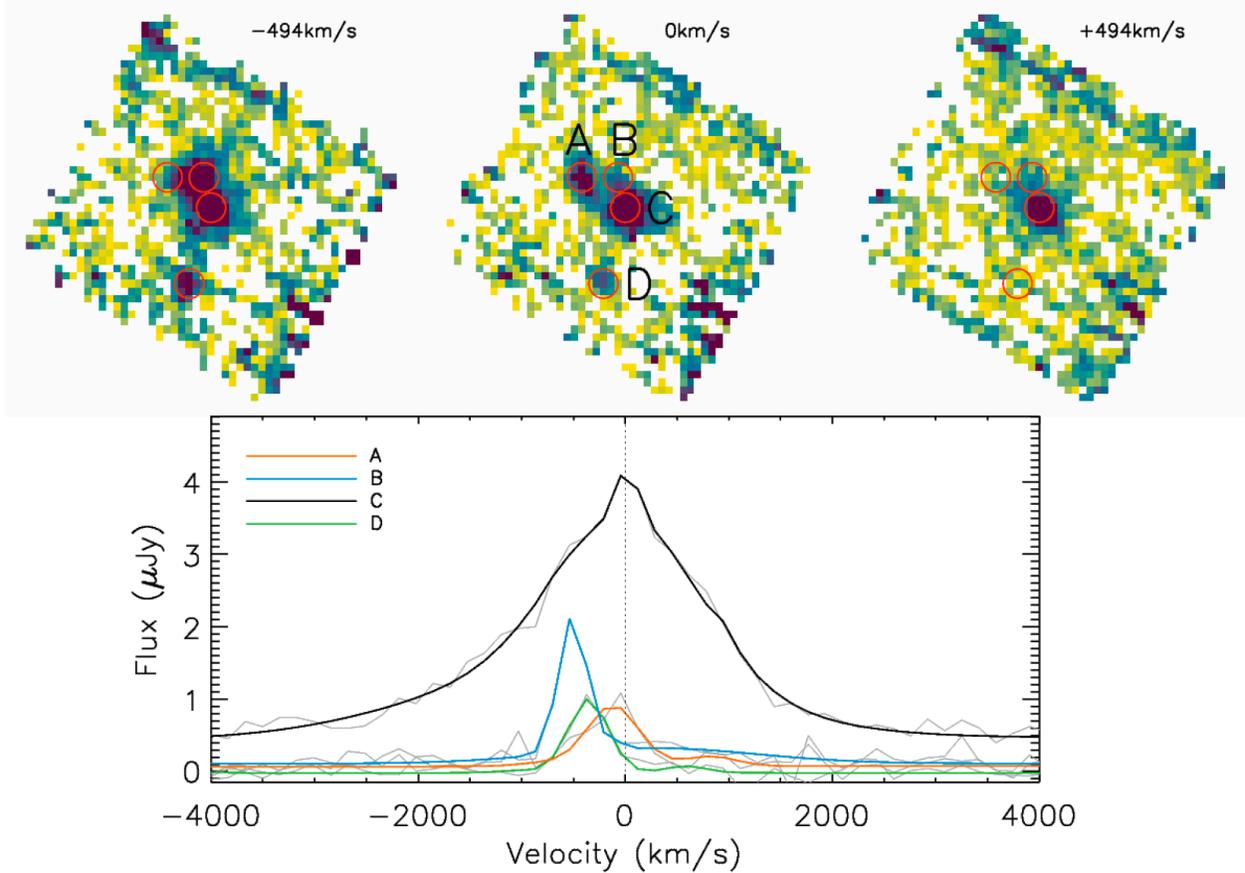

**Figure 3. JWST NIRSpec/IFU channel maps for the Hα emission line region.** Each map shows the Hα emission line fluxes in different velocity bins. The spatially extended outflow components A, B, and D are at velocity offsets of ∼ -600 – -500 km s$^{-1}$ with respect to the central broad-line component (C). In the bottom panel, NIRSpec spectra of each component are shown in the Hα emission line region. The spectra are extracted from circular apertures of radius of 0.2".

**Methods**

**1. Parent Sample**

The parent sample comprises a previously undiscovered population of black holes, identified as near-IR-dropout X-ray sources (i.e. invisible in the optical/near-IR bands) from the *Chandra*-COSMOS Legacy Survey[27,28], which consists of 4016 X-ray sources over a large area of ~2.2 deg$^2$. We used the multi-wavelength photometry from the most recent photometric catalog from COSMOS2020[46] and HELP[47], containing *GALEX* FUV, NUV, CFHT U, Subaru/HSC g, r, i, z, y, UltraVISTA Y, H, J, Ks, *Spitzer*/IRAC 3.6 μm, 4.5 μm, 5.8 μm, 8.0 μm, *Spitzer*/MIPS 24 μm, 70 μm, *Herschel*/PACS 100 μm, 160 μm, *Herschel*/SPIRE 250 μm, 350 μm, 500 μm photometry. We visually inspected all the optical/IR images and identified those without any optical counterparts within a 2" radius, which corresponds to the uncertainty of *Chandra* position. We excluded sources whose flux is contaminated by nearby bright sources and possible diffuse X-ray emission. This resulted in a final sample of 62 IR-dropout X-ray sources. All sources are detected in one or more of *Spitzer*/IRAC (3.6, 4.5, 5.8, 8.0 μm) bands and 26 sources are detected in *Spitzer*/MIPS 24μm photometry. 10 sources have *Herschel* far-IR detections. None of these sources has a counterpart in the VLA 3 GHz source catalog[48].

**2. ALMA Observations**

ALMA band 7 (275-373 GHz) continuum observations for all 62 IR-dropout X-ray sources were carried out in four observing blocks on 2019 November and 2022 January under the Cycle 7 program 2019.1.01275.S (PI: Suh) with a total of 42 to 46 antennas. The observations were centered on the *Chandra* X-ray positions with an integration time of ~5 minutes per source. The data reduction was performed using the standard ALMA pipeline version 2021.2.0.128 (CASA version 6.2.1.7). We measured the integrated flux of all our targets using the imfit procedure from the CASA pipeline. The sources were modeled with a circular Gaussian profile of variable total flux, centroid, width, axis and position angle. The 870 μm flux of LID-568 is 545+/-158 μJy, and the position of the 870 μm emission as measured from ALMA is in good agreement with those of *Spitzer*/IRAC. In Extended Data Figure 1, we show multi-band images of LID-568, which are invisible in the Subaru/HSC optical and UltraVISTA near-IR images.

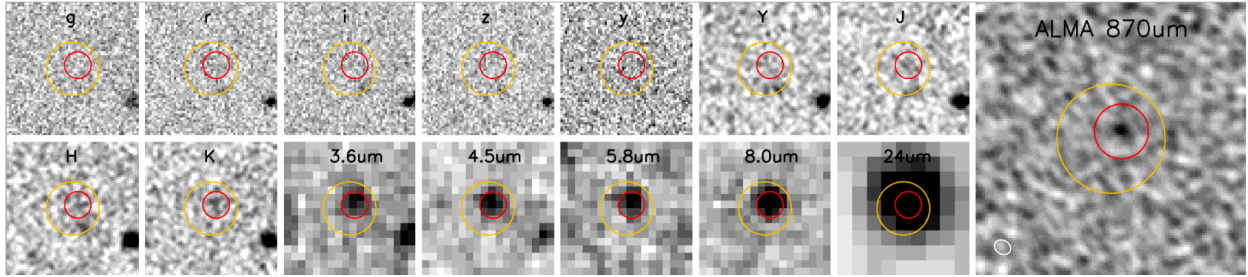

**Extended Data Figure 1. Multi-band images of LID-568** Multi-band 10"x10" images (Subaru/HSC *g, r, i, z, y,* UltraVISTA Y, J, H, Ks, *Spitzer*/IRAC 3.6, 4.5, 5.8, 8.0 μm, *Spitzer*/MIPS 24 μm, ALMA 870 μm) of LID-568. Yellow and red circles correspond to the *Chandra* X-ray positional uncertainty of 2 arcsec radius and the ALMA far-IR continuum position, respectively. The 5σ point source depth (AB magnitude) of the Subaru/HSC images are (g, r, i, z, y) = (28.4, 28.0, 27.7, 27.1, 26.6) and those of UltraVISTA are (Y, J, H, Ks) = (25.7, 25.4, 25.1, 24.9).

## 3. JWST observations

We obtained JWST/NIRSpec[49,50] and MIRI/LRS[51] observations of LID-568 under the Cycle 1 GTO program #1760 (PI: Suh). The NIRSpec/IFU observations were taken in April 2023 with the grating/filter combination of G395M/F290LP. This covers the spectral range of 3-5 μm with an average spectral resolution of R~1000. The field-of-view of the IFU mode is ~3"x3", with each spatial element in the resulting IFU data cube of 0.1"x0.1". We used the NRSIRS2 readout mode, which improves signal-to-noise and reduces data volume. The observations were taken with 18 groups and 1 integration per exposure, using a 4-point medium cycling dither pattern, resulting in a total exposure time of 1.45 hr.

The NIRSpec/IFU data reduction was performed with the JWST Science Calibration pipeline version 1.11.4, using the CRDS context jwst_1149.pmap. We also added additional steps in order to improve the resultant reduced data quality[52]. The reduction process consisted of three stages. The first stage accounted for detector-related issues, like bias and dark subtraction, and cosmic ray flagging. At the end of this stage, the groups were fitted to create 2D count rate images (i.e. "rate files"). The second stage applied the flat field correction, wavelength, and flux calibration. The calibrated exposures were then processed in the third stage, where a further flagging of cosmic rays was applied before building the final datacube. After running the second stage, we removed the 1/f noise affecting the rate files by subtracting from each spectral column its median value after applying a sigma clipping (see also [53,54,55]). We fixed a pipeline bug reported by the STScI Helpdesk by setting all the saturated pixels and the pixels with bad flat field correction to 'DO_NOT_USE', which removes several outliers from the calibrated exposures. We removed the remaining outliers from the datacube by filtering out all the voxels with a jump over contiguous channels persisting for less than four channels, which is the typical width of these features. Finally, we subtracted the background as a function of the wavelength by calculating the median over ten spectra extracted from empty regions in the cube field of view in each channel. We note that the background increases as a function of the wavelength, an expected effect due to an increase in the zodiacal and stray light[56]. We thus subtracted it channel-by-channel to obtain a background-free datacube.

The MIRI/LRS slit spectroscopy observations were conducted in January 2023 using a P750 disperser, covering a wavelength range of 5-12 μm with a spectral resolution of R~100. The observations were performed with 360 groups per integration in FAST/FULL mode, with 2 integrations per exposure using a 2-point dither along the slit. This resulted in a total exposure time of 1.1 hrs. The fully reduced data were retrieved from the Mikulski Archive for Space Telescope (MAST), which were processed using the JWST Science Calibration pipeline version 1.12.5, with the CRDS context jwst_1135.pmap.

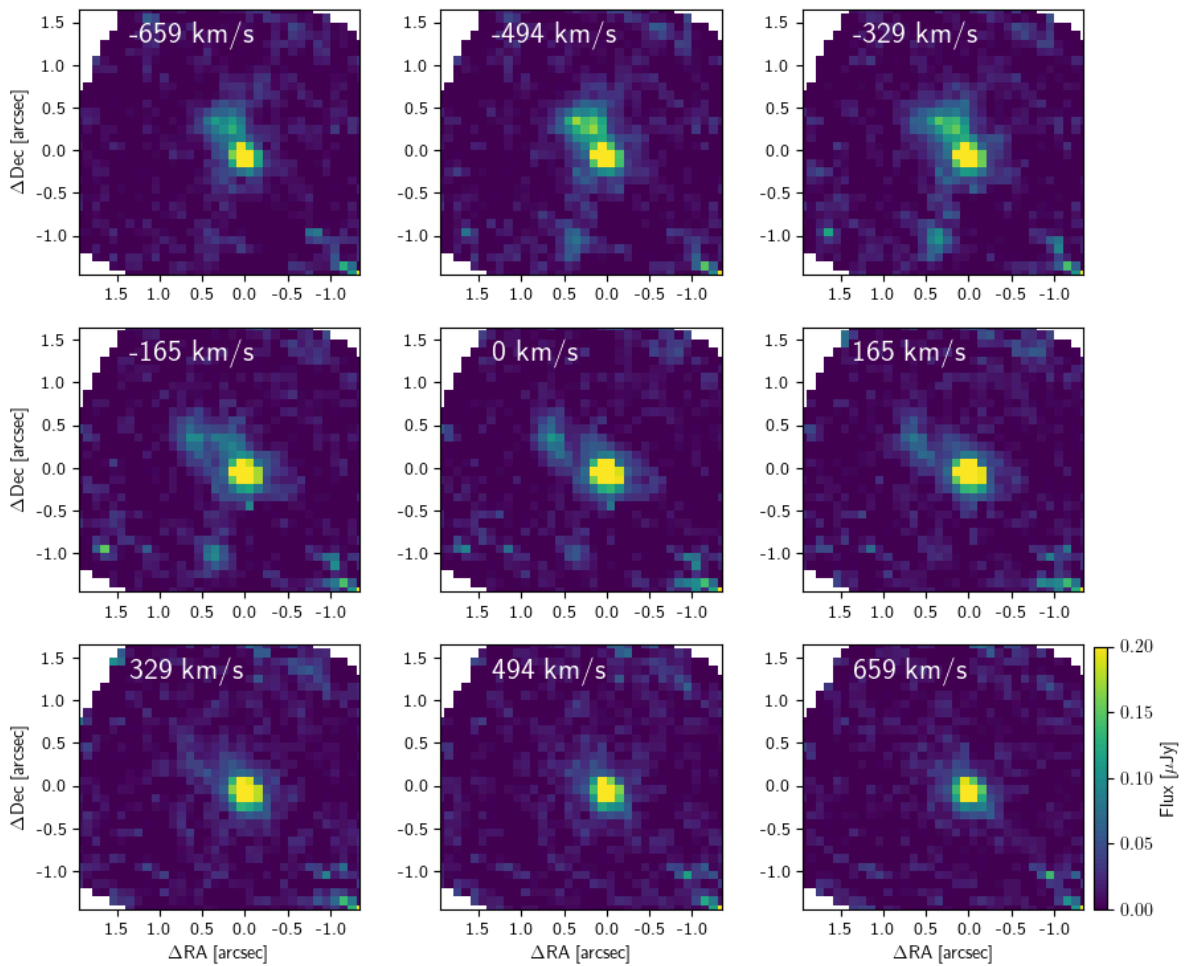

**Extended Data Figure 2. JWST NIRSpec/IFU channel maps for the Hα emission around LID-568.** Each map was created by averaging 3 neighboring channels. The maps are shown in single-channel steps corresponding to velocity steps of 165 km/s. The velocity marked in each map indicates the central velocity of the 3-channel average relative to the 0 km/s map centered at 3.259 μm. Spatial offsets in arcseconds are shown relative to the AGN location.

## 4. X-ray luminosity

To compute the intrinsic X-ray luminosity in the 2-10 keV band, we used the XSpec software (version 12.13.0)[57] to fit the *Chandra* spectrum using a simple power-law model with the photon index fixed to $\Gamma=1.9$, modified by both Galactic absorption ($N_H=2.6\times10^{20}$ cm$^{-2}$; ref. [58]) and absorption at the redshift of the source, $N_H(z)$. The second absorption component accounts for both nuclear absorption due to the gas orbiting in the proximity of the SMBH (i.e. torus) and absorption due to the interstellar medium in the host galaxy. The column density was measured as log $N_H=23.44$ [-0.34 +0.47], and the absorption-corrected, rest-frame 2-10 keV luminosity was determined as $\log(L_{2-10\ keV})=44.79$ [-0.33 +0.27].

To take into account the more complex absorption and reflection processes in the case of Compton-thick obscuration ($N_H>10^{24}$ cm$^{-2}$), we also derived the column density and intrinsic X-ray luminosity using the MYtorus model[59,60]. This model consists of three components: the obscuration along the line of sight, including Compton scattering, applied to the primary power-law, the reflection, and the fluorescence emission line complex. The relative strength of these components was fixed to be the same, and the inclination angle between the line of sight and the axis of the torus was set to 75° to ensure interception of the obscuring torus. A power-law photon index of $\Gamma=1.9$ was assumed. The column density and intrinsic X-ray luminosity derived from the MYtorus model were consistent with the standard power-law model, well in the Compton-thin regime. If we allow the photon index to be a free parameter, the fit tends toward a softer power-law ($\Gamma=2.4$ for MYtorus and 2.9 for the simple power-law) and, consequently, even higher column densities and intrinsic luminosities: log $L_{2-10\ keV}=45.08$ for MYtorus and 45.5 for the power-law. Therefore, the choice of $\Gamma=1.9$ was conservative in estimating intrinsic luminosity.

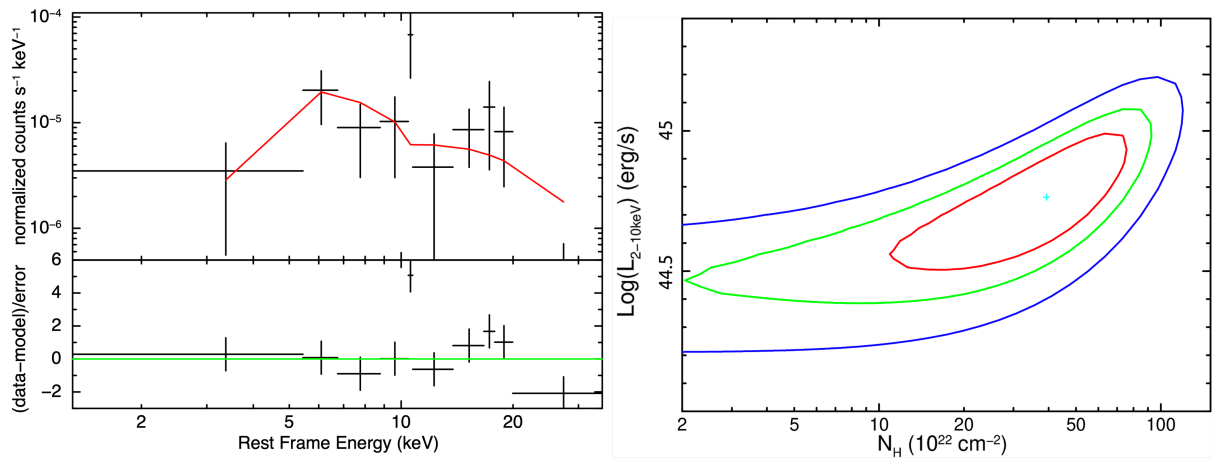

**Extended Data Figure 3. X-ray modeling** Left: rest-frame *Chandra* X-ray spectrum (left) and the data-to-model ratio (right) of LID-568. The best-fit model is plotted as a solid red line. Right: Confidence contours for two parameters of interest, column density and intrinsic luminosity. The red, green, and blue contours show the 68, 90, and 99% confidence levels, respectively.

## 5. SED fitting

We fitted the dust emission using the modified IR SED fitting code developed by [32]. We employed a composite mid-IR power-law and two-temperature greybodies. We used a fixed value of the emissivity ($\beta=1.5$), and allowed the mid-IR power-law slope ($\alpha$) as a free parameter. The rest-frame observed photometric data (black) are presented alongside the best-fit IR SED (yellow) in the left panel of Extended Data Figure 4. The SED is well-fitted by a power-law, hot greybody (655.5 K), and warm graybody (71.5 K) components, which are much hotter than what is typically observed in star-forming galaxies (10-60 K). From the best-fit, we derive the total IR luminosity ($L_{8-1000um}$) and the dust mass. In the right panel, we show the SED of LID-568 overlaid on the SED templates[61] of the AGN-dominated local ULIRG (Mrk 231), the extreme local starburst ULIRG (Arp 220), and the AGN dust torus model at redshift z=3.965. The IR SED shape of LID-568 seems to be consistent with the torus model spectrum, but cannot be explained by currently available IR SED templates of obscured AGN/ULIRGs.

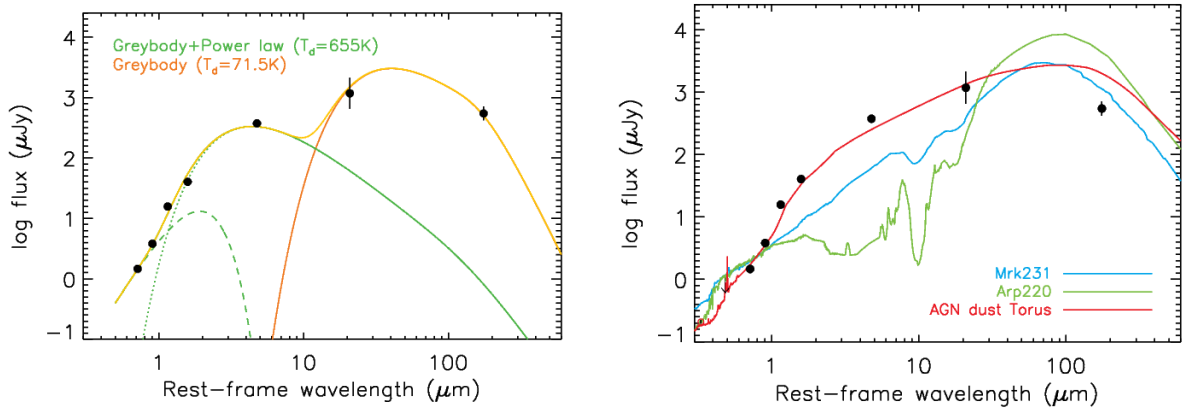

**Extended Data Figure 4. SED fit** Left: the rest-frame observed photometric data (black) with the best-fit model (yellow), including a power-law (green dashed), a hot greybody (655 K), and a warm greybody (71 K) component. Right: Overlay of the observed data (black) with the SED templates[61] of the AGN-dominated local ULIRG (Mrk 231), the extreme local starburst ULIRG (Arp 220), and the AGN dust torus model at redshift z=3.965.

## 6. AGN Bolometric luminosity

The bolometric luminosity of AGNs can be estimated from the X-ray luminosity by applying a suitable bolometric correction[62]. To accurately estimate the total intrinsic luminosity radiated by the AGN accretion disc, it is necessary to constrain the absorption-corrected intrinsic X-ray luminosity, as X-rays are often obscured and may include reprocessed radiation. The AGN bolometric luminosity of $L_{bol}=46.59$ erg s$^{-1}$ is derived using the absorption-corrected rest-frame 2-10 keV luminosity by applying a luminosity-dependent bolometric correction described in [62].

We also computed the AGN luminosity from the SED by integrating absorption-corrected total X-ray luminosity ($L_{0.1-100keV}$) and the best-fit AGN torus luminosity ($L_{1-1000um}$) following [63]. To convert the IR luminosity into a proxy for the intrinsic nuclear luminosity, we consider the geometry of the torus and its orientation by applying the following correction factors: the first correction is related to the covering factor, which represents the fraction of the primary UV-optical radiation intercepted by the torus (~1.5; ref. [64]), and the second correction is due to the anisotropy of the IR dust emission, which is a function of the viewing angle (~1.3; ref. [65]). The derived AGN bolometric luminosity is $L_{bol}$=46.68 erg s$^{-1}$, which is consistent with the X-ray-derived bolometric luminosity.

We additionally derived the bolometric luminosity using the H$\alpha$ luminosity following [3]. We calculated the rest-frame 5100Å luminosity from H$\alpha$ luminosity using equation in [66]. The bolometric luminosity is estimated using the bolometric correction factor in [67], $L_{bol}$=10.33x$L_{5100}$, is 45.60 erg s$^{-1}$, ~1 dex lower than that derived from other methods (i.e. X-ray luminosity and SED-fitting). This indicates that the H$\alpha$ emission could possibly be highly obscured, potentially leading to an underestimation of the H$\alpha$-derived black hole mass by a factor of a few. However, we point out that when estimating the Eddington ratio using an internally consistent method based on the H$\alpha$ emission for both AGN bolometric luminosity and the black hole mass, the black hole is still accreting at the super-Eddington accretion level of ~4.4. We note that the bolometric correction factor for H$\alpha$ luminosity could be uncertain for those obscured AGNs with a super-Eddington accretion at high redshifts.

The estimated bolometric luminosities are shown in Extended Data Figure 5.

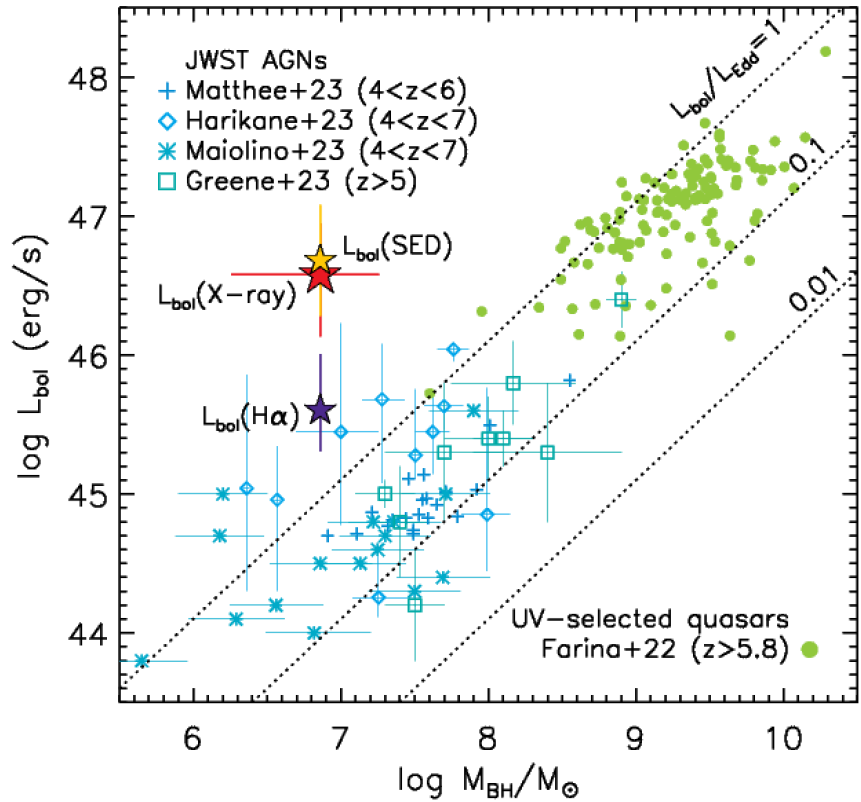

**Extended Data Figure 5. AGN bolometric luminosity ($L_{bol}$) versus black hole mass ($M_{BH}$) of AGNs at high redshift.** The AGN bolometric luminosities of LID-568 derived from X-ray (red star), SED-fitting (yellow star), and H$\alpha$ luminosity (purple star) are indicated.

## 7. Black hole mass and outflows

The single-epoch virial black hole mass is estimated using the broad H$\alpha$ emission line width and the line luminosity from the rest-frame UV/optical spectra as a proxy for the characteristic velocity and the size of the BLR. The NIRSpec spectra were extracted from a circular aperture centered at the position of the broad-line region, with radius of 0.2" (r=2 pix). We utilized the mpfit routine for fitting the emission lines, employing a Levenberg-Marquardt least-squares minimization algorithm to derive the best-fit parameters and assess the overall fit quality (refer to [68]). Specifically, we fitted and subtracted a power-law continuum, $f_\lambda \propto \lambda^{-a}$, from the spectra, and fitted the narrow emission lines of [NII]6548, 6583 Å with a fixed ratio of 2.96, along with H$\alpha$ 6563 Å. The broad H$\alpha$ line was fitted with two broad Gaussian components: one for the broad-line region and the other for the blue-shifted outflow component. Finally, we measured the broad-line width and the line luminosity from the best-fit spectra. The black hole mass was computed using the equation from [66]. While the measurement uncertainties on $M_{BH}$ are relatively small (~0.1 dex), systematic uncertainties associated with different single-epoch virial calibrations carry a scatter of ~0.3 dex[35,36,69]. We estimate the black hole mass to be 7.2 [-5.4 +10.8]×$10^6$ $M_\odot$. The uncertainties of the black hole mass are determined by the sum of the statistical and intrinsic scatter of the calibrations.

Extended Data Figure 6 (left) shows the best-fit model around the H$\alpha$, [NII], and [SII] region. Broadened and/or shifted components in emission lines trace gas with different kinematics, potentially indicating outflows. We investigate possible signs of outflows using H$\alpha$ and [SII] lines since [OIII], which typically serves as a tracer of outflows, is not covered by our dataset. In Extended data Figure 6 (right), we compare the blue-shifted H$\alpha$ emission line profile with that of the [SII] line. The broad blue-shifted emission is evident in both the H$\alpha$ and [SII] lines, exhibiting the same broad line width. From the best-fit model, we infer a spatially unresolved outflow velocity of ~ -540 km s$^{-1}$. Similar outflow velocities are associated with the spatially extended H$\alpha$ emission (see Extended Data Figure 2). The mass of the ionized outflow as inferred from the blue-shifted outflow component of the broad H$\alpha$ emission is 1.4x$10^7$ $M_\odot$, using equation (1) from [70]. Assuming an outflow velocity of -540 km/s and $r_{out}$=1" (~7 kpc; the spatial extent of H$\alpha$ features), the outflow rate is ~ 3.1 $M_\odot$ yr$^{-1}$.

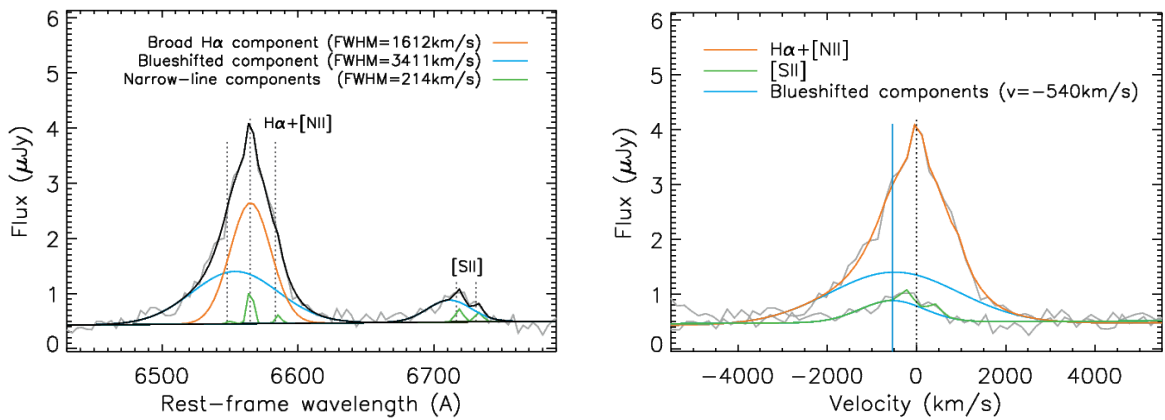

**Extended Data Figure 6. Hα broad-line fitting** Left: The JWST NIRSpec spectrum (gray) with the best-fit model (black). The spectrum is extracted from a circular aperture of radius 0.2" centered in the central broad-line region. The power-law continuum, narrow-line components (green), broad-line components (orange), and outflow components (blue) are indicated. Right: Comparison of the blue-shifted (outflow) line profiles of the Hα+[NII] and [SII] in velocity space. The blue-shifted (outflow) components are observed at a velocity of ~ -540 km/s relative to systemic.

## 8. Environment

We measure the environmental density surrounding LID-568 by employing the Voronoi tessellation Monte Carlo (VMC) mapping described in [71,72]. Briefly, this technique uses a weighted combination of spectroscopic and photometric redshifts to construct a galaxy overdensity cube in thin (7.5 proper Mpc) slices running from $2 < z < 5$. The mapping leverages the wealth of panchromatic imaging data in the COSMOS, as well as a large number of spectroscopic redshifts drawn from public surveys as well as proprietary data. The particular instance of the VMC mapping used in this work is identical to that of [73].

Once an overdensity cube is constructed over the full redshift range, a Source Extractor-based post-processing technique described in [59] is used to link detections of overdensities across contiguous slices to search for coherent structure and estimate the mass of detected structure. For density mapping at z>2, this post-processing technique is trained on mock observations of custom-built light cones in order to maximize the purity and completeness associated with the detection of protogroups and protoclusters. At the spatial location of LID-568, the spectroscopic coverage in the COSMOS field is fairly sparse, and we estimate that our method is only >50% complete for structures with z=0 masses greater than $10^{14.5} M_\odot$, i.e. massive protoclusters, at z~4.

We find no evidence that LID-568 is associated with an overdensity of galaxies. The local overdensity at the location of LID-568 is log $(1+\delta_{gal})$=0.11, which is approximately a 1σ fluctuation over the mean (galaxy) density of the universe at these redshifts. Measuring the average overdensity in a cylindrical aperture of radius 1 proper Mpc and depth of Δz=0.02 centered on the redshift of LID-568 recovers a consistent value. Additionally we detect no associated coherent structure within Δz=0.04 and R<5 proper Mpc of LID-568, which indicates it is likely not embedded in a massive forming cluster. However, given the paucity of spectroscopic redshifts in proximity to LID-568, we cannot rule out membership in a lower mass structure. We also note that, at such redshifts, galaxy-traced methods can fail to detect massive overdensities that are well traced by neutral hydrogen[74]. Future spectroscopic observations of the surroundings of LID-568 and similar sources will help to better quantify the environments in which they reside.

**Methods References**